# The Current Chinese Global Supply Chain Monopoly and the Covid-19 Pandemic

George R. Rapciewicz, Jr., D.B.A. Candidate

Donald L. Buresh, Ph.D., J.D., LL.M.

**Abstract**

Because of the ongoing Covid-19 crisis, supply chain management performance seems to be struggling. The purpose of this paper is to examine a variety of critical factors related to the application of contingency theory to determine its feasibility in preventing future supply chain bottlenecks. The study reviewed current online news reports, previous research on contingency theory, as well as strategic and structural contingency theories. This paper also systematically reviewed several global supply chain management and strategic decision-making studies in an effort to promote a new strategy. The findings indicated that the need for mass production of products within the United States, as well as within trading partners, is necessary to prevent additional Covid-19 related supply chain gaps. The paper noted that in many instances, the United States has become dependent on foreign products, where the prevention of future supply chain gaps requires the United States restore its manufacturing prowess.

## Introduction

The purpose of this paper is to provide an accurate assessment of the current state of global supply chain management due to the Covid-19 pandemic. At issue is whether supply chain management organizations within the United States are dependent upon Chinese made products, whether the United States-based supply chain management organizations need to implement additional strategies to prevent future supply chain shortages, and whether the United States should increase production internally to avoid future supply chain shortages. This essay develops the notion that global supply chain management ("GSCM") is a contingency theory candidate. The paper examines existing literature, including online sources, to determine whether contingency theory is a critical factor in sustaining future GSCM systems. The article also focuses on the impact of uncertainty theory when evaluating GSCM.

## The Current State of the Literature

The purpose of this literature review is to describe how Covid-19 has influenced the outcomes from supply chain management systems. Juan and Lin focused on the effect of Covid-19 regarding United States trade policy and the disease's impact on the global supply chain management system [1]. The authors identified that some supply chain organizations would recover faster than others using the supply chain resilience hypothesis [2]. It is probably still too early to understand the long-term impact the past events will have on GSCM. Before the current viral pandemic, the majority of GSCM was monopolized by the Chinese manufacturing industry. This was prevalent given the shortage of medical masks, hand sanitizer, and the ability to obtain ventilator systems to support the continued increase of individuals falling ill due to the virus. This paper will discuss the ongoing situation with GSCM and provide a suggested method to deter such incidents from happening in the future. Scholarly research will be utilized to support arguments being made to discontinue dependence upon foreign entities that produce products for American citizens.

Haren and Simchi-Levi predicted that in mid-March 2020, the effects of Covid-19 on GSCM would force thousands of companies to shut down due to the potential for the spread of the virus, thus disrupting the GSCM's ability to deliver products over 30 days [3]. Before Covid-19, Severe Acute Respiratory Syndrome ("SARS"), which took place during 2002 and 2003, the effects on the market were so minuscule that no one recognized any impact at all [4]. At that time, China only controlled just over four percent of the world's gross domestic product ("GDP") [5]. Today they control over 16 percent of the world's GDP. When China quarantined more than half of its population, this caused all manufacturing of products to come to a complete halt and disrupted the GSCM in a manner not seen since the first world war [6]. People wanted more cost-effective means of obtaining products, and manufacturing within countries such as the United States and Europe was not beneficial due to higher costs, thus resorting to these countries to rely on China for decades. Businesses want to make products more cost-effective manner to increase sales and profits. Ironically, America and Europe are now manufacturing products within their own countries, which is resulting in work stoppages in China.

The need to provide a contingency plan to prevent future disruptions of the supply chain management due to current supply chain management system requirements for collaboration and the effects of visibility, velocity and flexibility when cooperation is disrupted is prevalent [7]. Panjiva Research identified that the automotive, company goods and electronics industries were immediately impacted during Covid-19 [8]. Tangpong et al. observed that the contingency theory can directly influence executive leadership abilities to make appropriate decisions [9]. Flynn et al. noted that the need to accept uncertainty based on classical organization theory and information processing theory, due to their apparent presence within supply chain organizations [10]. Uncertainty is commonplace within the world and supply chain organizations, and managing uncertainty is something that organizations must do.

Classic organization theory can be influenced by an organization's independence from other organizations, where the culture, functional design, existing resources, and their current motivations for conducting business are different [11]. Uncertainty, as it relates to the contingency theory, is dependent on the form, fit, and function of the organization's processes, structure, and environment [12]. When looking at customer satisfaction, customer base, current suppliers, and competition, topped with extensive regulatory requirements, these factors can vary based upon the country of origin [13]. Uncertainty consists of physical manifestations, behavioral response, perception, and social expectations [14].

Hahn and Popan defined how an individual's behavior can influence their intentions, and how the theory of reasoned action ("TRA") can be used to determine behavioral intent before it happens [15]. With uncertainty, TRA may be a useful tool to identify behavioral intention and potentially alleviate some, if not all, of its associated uncertainties. On its face, uncertain behavior appears to be the main component. However, it would seem that the opportunity to extend TRA using those four factors could provide statistically significant results.

Goldsby et al. showed that organizations in the supply chain industry do not keep current with planning and operational processes, tending to take action only when a social, political, economic, or environmental crisis arises [16]. The results from Flynn et al. seemed to be related to contingency theory, taking into account population growth and immigration [17]. Global supply chain organizations are faced with a dilemma. Every country has its own unique culture, where even in the United States, different regions of the nation demand different kinds of goods and services. The question is, how can TRA fit into the contingency theory when culture may affect the variables necessary to identify behavior.

The theory of planned behavior ("TPB") is another methodology that can be utilized to measure behavior. Unlike TRA, which is used to identify behavioral intent, TPB is employed to prevent a specific action [18]. It seems that a model that blends TRA and TPB with contingency theory warrants an investigation. Mandal observed that behavior as an inherent risk associated with supply chain management encouraged organizations to exercise flexibility, thereby ensuring risk management practices are affirmed [19]. However, with cultural differences in a global market, risk management becomes paramount. Mandal demonstrated that technology is an additional factor in managing risk [20].

The technology acceptance model was created in 1986 by Fred Davis and was derived from TRA [21].

Technology is now an additional factor to consider in a behavioral model. Mandal (2016) expounded on supply flexibility, where the term "agility" is defined as an organization's ability to adapt to a diversified environment, to overcome business-related obstacles, and to ensure the organization's continued existence. Martin noticed that with agile supply chain management ("ASCM") theory, organizations should consistently be able to improve processes while meeting supply and demand requirements [22].

Maurer took a different approach to supply chain management, where he focused on the economic impacts that drive supplier diversity [23]. He pointed out that self-governance in supply chain industries relies on coercion and does not influence violence inside the organization based upon an organization's profit and loss [24]. However, although this approach is viable, it is not entirely accurate. Organizations experienced serious conflict both inside and outside during the stock market crash of 1929 and the Enron scandal in 2001 [25]. The morale of Maurer's research is that with great power comes great responsibility [26]. In a global market, supply chain management, organizational failures can affect other organizations, that trickle eventually down to the consumer. Politics can play a crucial role in not only ensuring consumers are protected, but could also force organizations to close their doors [27]. For example, when Rep. Alexandria Ocasio-Cortez and other New York state and city politicians forced Amazon to rethink its strategic move into New York City, the company relocated to another state [28].

Like Mandal's [29] findings, Roy observed that an organization's agility regarding rapid strategy implementation, but is more focused on supply and demand in forecasting future potential profits [30]. Roy argued that forecasting could be a labor-intensive activity when attempting to achieve positive results [31]. Roy focused on consumers and their use of technology to find the best-priced product [32]. By employing data analytics to identify logistical e-commerce business models that specialize in supply chain management may help clarify what technologies users are utilizing effectively [33].

Wei focused on corporate political activities in China, showing significant relationships between critical variables associated with the benefits of political activities and business [34]. The study utilized a focus on resource dependency theory ("RDT"), with 78 variables identified using a 5-point Likert-like scale [35]. Wei sent out 350 questionnaires and obtained 233 responses [36]. Only 201 responses were deemed valid, resulting in a 57.4 percent response rate [37]. The result of the reliability and validity verification process was that five of the 78 variables were inadmissible [38]. Wei reported that there was a positive statistically significant relationship at the 95 percent confidence level existed between corporate political strategies and corporate political resources [39].

Wei noticed that government involvement influenced intangible and relational resources [40]. Direct participation was both positively related to organizational and relational resources [41]. There was also a positive and significant relationship among all of the variables tested during the regression analysis [42]. The study concluded that political activities in China benefited corporations as a whole by influencing government decision-makers but resulted in ethical, moral, corruptive, and illegitimate practices being conducted [43]. Financial incentives, such as bribery and gaining political influence, were the critical parts of political activities [44]. The main goal of organizations identified by Wei was to have a strong affiliation with key political entities [45]. Part of the Chinese corporate strategy is to intermingle itself with government officials to encourage economic growth, increased profits, and organizational resource expansion [46].

**The Gaps in the Literature**

Some authors have not considered the human behavioral factors that could influence risk, uncertainty, and progress. Most of the studies above were geared towards business operations and planning strategies, such as risk mitigation, process monitoring, and improvement. Additional customer engagement factors were not associated with behavioral variables that could be used to identify appropriate strategies in sustaining a diversified global environment.

After reviewing 113 articles, Lebek et al. found that of 54 theories implemented, the theory of planned behavior ("TPB"), general deterrence theory ("GDT"), protection motivation theory ("PMT"), and technology acceptance model ("TAM") was primarily used [47]. The data were derived from ten years of research on information security and information systems [48]. Supply chain management organizations depended on technology even more than on their suppliers. It seems that it is critical to incorporate behavior into every supply chain management process, regardless of the uncertainty variables such as physical manifestations, behavioral response, perception, and social expectations, and how they will impact an organization's supply chain system. Even when an organization implements all of the control measures they believe are necessary, if it is unknown how the predicted behavior interacts with sustainment processes, no risk management plan, forecasting, contingency plan, or supply flexibility methodology will be adequate.

**A Little Background Information**

An understanding of both GSCM and economics is necessary to understand the current situation, and how the Covid-19 pandemic is affecting the global economic framework. Manufacturing creates jobs that generate profits and tax revenues and fuels the economy. A centralized production model or node has the potential to create a single point of failure, which can hinder economic development [49]. A quick review of China's role is in order to develop a better understanding. Previous scholars predicted that China would become a manufacturing behemoth, impacting the world in good times and bad. Kremer's O-Ring theory ("ORT") indicated that manufacturing techniques would require more technology due to specializing in complex products such as airplanes [50]. This capital investment increased costs in the short-run due to implementing high-end technologies to streamline processes. Based on O-Ring theory, prices increased because highly-skilled people were needed to operate the technology [51]. For example, dentists use technology that was not available 20 years ago, which has drastically reduced moderate periodontal disease. Laser technology cauterizes the gums to reduce receding gums and prevent tooth loss [52].

TAM is utilized to determine whether to accept or reject a particular technology before implementation. TAM was created in 1986 and derived from the theory of reasoned action ("TRA") [53-55]. Society has come to accept technology regardless of its outcome. GSCM depends upon technology to ensure its success, but it may also be a hindrance. The United States is increasingly dependent on technology because the nation relies on other countries to manufacture quality products at low costs. For example, SAP, AG requires a server, an internet connection or local area connection, a network administrator, a system administrator, a cybersecurity implementer, as well as users, or at least six individuals just to operate one system. Taj observed that 65 Chinese-based manufacturers identified nine variables associated with running an entire manufacturing plant; inventory, team approach, processes, maintenance, layout/handling, suppliers, setups, quality, and scheduling/control [56].

There are many subcategories within the nine variables present in a manufacturing plant. Taj found that Lean processes were not employed because current Chinese laws prevented lowering labor costs [57]. Additionally, Taj observed that Lean processes would benefit Chinese manufacturing by increasing not only quality levels but also decreasing consumer costs [58]. This could be a death sentence to Chinese manufacturing organizations, where quality is sometimes sacrificed for quantity and cost. Out of all of the items produced by China, petroleum scored the highest when considering automation [59]. Still, petroleum-based products are generally expensive worldwide, so it seems that automation costs could easily be absorbed into acceptable expenses associated with purchasing the product.

The second-highest scoring product produced by China is telecommunications-related products, which a vast majority of U.S. customers purchase. Out of a scoring system of 100, the Chinese average manufacturing quality score was fifty-five [60]. This might explain why products are so cheap when made in China not only on a cost basis but on a quality basis. BestBuy, which is one of the largest retailers of electronics in the United States, purchases goods from 170 Chinese manufacturers [61]. Quite reasonably, BestBuy's stock has nearly doubled since 2015 by purchasing products from foreign manufacturers [62]. The Trade Agreement Act ("TAA") ensures that the majority of products manufactured within the United States or by approved countries abide by American manufacturing standards [63].

**Should Existing Theory be Revisited?**

Contingency theory, classical organization theory, and information processing theory exist within supply chain organizations because they are all focused on uncertainty [64]. Have U.S. and European countries and manufacturers looked into these theories? Other ideas to be applied are TRA and the theory of planned behavior ("TBD") [65]. When trying to mitigate uncertainty, TRA would identify behavioral intent and possibly alleviate some, if not all, of the potential tensions regarding customer satisfaction, customer base, current suppliers, and competition that drive an organization's goals [66].

The theory of planned behavior ("TPB") is another theory to consider. TPB can be used to define controls for a particular behavior to prevent a specific follow on action [67]. The TAM can also be utilized within GSCM practices by taking into account all of the variables associated with technology. Implementation of TAM can ensure that the human factor is accounted for not only at the user level but also at the organization as a whole, both internally and externally. Given the current world situation and China's monopolization of GSCM, buying a simple system that requires multiple users and administrators to sustain it may not be feasible. Most organizations are looking at cost reductions. It should also be noted that China has a history of stealing American corporate trade secrets, technologies via espionage [68]. What is not taken into account is uncertainty in a broader spectrum. According to Biggs and Buchler, perceived usefulness, perceived ease of use, behavioral intent, and attitude towards use could be extended within the TAM using TPB and TRA variables [69]. Additionally, Practice-Based Research ("PBR") could also be developed into TAM [70]. Aside from traditional research methodologies, PBR adds four additional variables identified as the role of text and image, relationship of form and content, the function of rhetoric, and function of experience [71].

Why look into these suggestions? To prevent a future scenario where the United States is completely stifled in providing products to sick individuals, while simultaneously hurting the economy, could create mass hysteria because of one country's ability to control the vast majority of a market. This is why globalism is both dangerous and profitable. The U.S. has not felt the trickle-down effect, in which one action creates a wave of additional activities that can affect things in its path [72]. A prime example of this is the recent events in the U.S. Senate, where funding is attempting to be passed. Still, on both sides of the aisle, politics continues to play out, adversely affecting the U.S. public. The exploitation of people suffering and politicians' pet projects trying to be slipped into bills that are not even relevant to the current crisis and proposed funding are a daily activity [73]. Finger-pointing is abundant, while people continue to struggle to make their mortgage and rent payments [74]. Food and other necessities may not be able to be purchased due to a lack of jobs and money to buy them [75]. This may be the tip of the iceberg, where this experience could be used as an example to change the way the U.S. public operates.

Hessels and Terjesen wrote about the hypothetical scenario where resource dependency and institutional theory determine a manager's decision whether to export or not to export products, as well if the decision to ship to an indirect or direct customer is most appropriate [76]. It was a quantitative study that sought to test four hypotheses utilizing binominal regression analysis techniques based upon empirical data [77]. The independent variables are export involvement (no export activity and exports, both direct and indirect), as well as export mode (direct export and indirect export) [78]. Knasko demonstrated that there could be more than one dependent variable due to potential carryover effects [79].

Knasko utilized both foreign and domestic suppliers and distributors with a sample size of 402 for the binominal regression analysis for export involvement [80]. A sample size of 118 was obtained for the export mode binominal regression analysis [81]. Knasko concluded that institutional theory perspectives influenced managers' decisions to export, where resource dependency theory influenced the decision to utilize a direct or indirect export mode [82]. Hessels and Terjesen identified that support organizations in-between supply chain management organizations such as their suppliers and distribution centers, should further examine their roles to determine if outside influence could affect export mode and export decision making [83]. It should be noted that resource dependency and institutional theories are not the same when it comes to the prediction of an organization's performance [84].

**Industries that are Impacted Positively or Negatively**

If this topic were discussed back in January 2020, the research may have been quite different. The only similarity that would have been identified is that China owns and controls the GSCM system. General Motors is currently manufacturing cars, trucks, as well as ventilators to save lives. Fiat, another automobile manufacturer, is making surgical masks to deter infections [85]. Hospitals are struggling to treat patients due to a lack of beds and their staff getting sick, as well as a shortage of Covid-19 testing kits, and pretty soon antibiotics [86]. Law enforcement is getting ill, and judges are releasing prisoners due to fear of them dying in prison, and our criminal justice system is being adversely affected [87]. The sectors involved are law enforcement, the justice system, automobile production, and healthcare [88]. How are other sectors being affected? America's national schooling system for grades K-12 [89]. There is even a discussion of children's school's graduating kids several months early due to the coronavirus [90].

All of the above is affected by the trickle-down effect. As more and more prisoners are released, and supplies become scarcer, there will be an uptick in criminal activity [91]. With the current civil unrest, the supply chain management within the United States is further tested [92]. Because there is not enough law enforcement in place, and judges releasing criminals due to Covid-19, it is expected that repeat offenders will offend again [93]. If President Trump decides to lift the stay at home order due to economic concerns, then more people may die [94]. There has been an increase in infections, possibly due to the lack of social distancing [95]. If the government decides that the economy is more important than social distancing, only time will tell if the supply chain management system and the economy will bounce back [96]. One can only hope that the spread of infection does not increase sickness, thus causing even more bottlenecks in the GSCM system.

It seems that Hong Kong's relaxed attitude towards Covid-19 and the reopening of the city erased the gains it earned. [97] In September 2019, NBC News reports Breslauer and Dilanian conducted an investigative report, in which they identified that China controls the majority of the production of antibiotics throughout the world. United States national security advisors noted that if China stopped providing these life-saving drugs from the U.S., all hospitals would cease to exist within months [98]. By controlling the GSCM system, China may be able to destroy our way of life without even firing a single shot.

**Mitigating a Negative Impact**

It appears that the United States must discontinue its dependency on foreign products by controlling its vital resources and manufacturing efforts. If the current situation is evaluated and lessons learned synopsis is created, the United States can develop an appropriate contingency plan based upon known variables. Not uncertainty. The U.S. could then further evaluate uncertainty and develop additional contingency plans to ensure our supply chain management system is sustainable should another event occur. America could continue to research hybrid automobile technologies, enhance shale oil and fracking processes, and become less dependent on other nations to support our needs [99]. There would be a reverse trickle effect in the U.S. economy, adversely affecting the Chinese economy [100]. So, should the United States continue to support globalism, knowing that if another virus hits the homeland? The U.S. economy and its citizens are at risk, with the possibility that the country may not survive a second wave [101]. The United States needs to bring the production of life-saving drugs and materials back to the

United States as soon as possible [102]. The country needs to renegotiate with nations that are not only friendly to us but also friendly to the environment with regards to GSCM and trade [103].

The behavior of the U.S. public and how that behavior relates to the decisions made in supply chain management organizations are not being adequately incorporated into the identified strategies. According to Saunders et al., there are three forms of usage of a literature review: preliminary search, critical review, and proper placement of the research findings [104]. Supply chain sustainment utilizes these methodologies to continue ongoing operations [105].

What this article has identified is the lack of measuring of behavior and incorporating controls to meet standard supply chain management practices for ongoing operations and forecasting events. The necessity to incorporate appropriate behavioral measurement tools as it relates to technology has been identified is inherent in the literature above. As demonstrated by Angonga and Florah, the ability to provide a different perspective on human behavior, how it can influence current theories, and allow organizations to not focus primarily on a set of processes while discounting other methodologies is critical [106]. This study has identified some of the resources, indicating that supply chain management is focused mainly on strategies, but does not take into account other variables related to technology and behavior. The usage of the strategic contingency theory could impact future GSCM operations by alleviating supply chain management stoppages [107].

**More Ways to Improve Global Supply Chain Management**

Additional research in creating a production environment within the U.S. that is efficient and environmentally friendly is necessary for America to become less dependent upon Chinese made products. Further research is needed to extend the TAM model to support supply chain management practices that implement technologies that do not create additional labor costs associated with multiple user interfaces. Further research on behavioral intent with regards to TRA, TPB, and PBR methodologies as America progresses in the implementation of stateside production is warranted. This is necessary as human behavior has become the main focus during the stay at home orders [108]. Additionally, ASCM seems to be sufficiently flexible to meet supply and demand practices [109]. Economic and political factors that may influence the implementation of a massive stateside production environment need to be considered [110]. Finally, additional research on environmentally friendly fracking processes may be warranted.

Organizations that utilize resource dependency theory ("RDT") and transaction cost theory ("TCT") to determine whether the decision to outsource information technology services that support university libraries that are open to the public and located in Kenya were identified vital issues [111]. RDT focuses on resources that identify the resources are the lifeblood of an organization, where the country becomes dependent upon to sustain its business operations [112]. TCT dictates when a firm can perform specific tasks, other tasks can be performed directly related to the market, determining whether or not to in-source or outsource resources [113]. The study examined multiple services associated with information technology, such as helpdesk support, printing services, and web design, to name a few [114].

One significant finding was that 50 percent of the staff of a company that required a particular service was not involved in the vendor selection [115]. The team ended up receiving a vendor that provided a system that did not meet their requirements, and work stoppages began [116]. This is intriguing based upon the intent to extend the TAM model to incorporate PBR, which, in theory, could ensure that implementation of technology without consulting all parties involved could have significant negative impacts on productivity. This mirrors a subjective norm, which means that a senior leader or manager can influence the decision to go with an unreliable vendor, against the wishes of those who desperately need the system [117]. Even in the case where the staff was not involved in the decision to hire a particular vendor, the subjective norm is still prevalent, where key stakeholders were not included in the selection process [118]. Mwai et al. concluded that RDT performed well as long as organizations looked beyond current constraints and ensured that the organization did not become dependent upon their vendors [119]. TCT findings were, in fact, very similar to RDT [120]. In contrast, caution should be exercised before the implementation of outsourced resources due to increased administrative costs, as it was suggested the in-sourcing would be the most feasible method to sustain the library [121].

**Redirecting Global Supply Chain Management**

China generates the most pollution in the entire world [122]. This is due to the country's heavy manufacturing processes and nearly zero environmental controls [123]. Aside from the quality of their products and lack of transparency, China will most likely not change their methods of production any time soon unless forced to do so and will continue to control the market [124]. This appears to be the only way to force China is by bringing back manufacturing jobs to the United States [125]. McFarlane suggested that the school of thought theory could be based upon the notion that religious ideology, related politics, and geographical locations may determine potential peculiar political views within the United States, and how these perceptions influence the implementation of that philosophy [126]. It will probably take several years to create a production environment back in the United States, but the U.S. can learn from China's mistakes [127]. The United States is already at the forefront of being an environmental leader [128].

By starting this process now, the United States can identify the gaps in the implementation of mass production within this country. Most supply chain management studies are driven towards understanding and defining operations and planning strategies associated with SCM, and continuously overlooking human factors and behavioral intent [129]. The bottlenecks are utilizing practices such as ASCM and Lean processes, while further improving our environmental impact while analyzing behavioral intention, where potential variables can be measured. These are not only life-threatening measures the U.S. must take but also other real business problems that need to be overcome. The Covid-19 will be here for years, say, scientists, researchers, and business and economic experts [130]. America has the opportunity to change and create its own sustainable way of life without depending on other nations [131]. Imagine if China decided to withhold all antibiotics from the United States right now. The effects of such a policy could have dire national security implications, possibly leading to war between the United States and China.

One could utilize political, geographical locations, and even religious ideology in determining the differentials related to the political views that are prevalent of each of the fifty states [132]. According to Dragomir, no company or organization can grow or sustain its operations without using management methodologies, rules, and regulations, and well as the appropriate apparatuses necessary in a diversified and complicated environment such as those in the supply chain management [133]. Dragomir identified how behaviors must be incorporated into business management processes [134]. Defining and measuring the behaviors also stimulates an organization's internal and external procedures, thus effectively increasing the success rate of proposed business goals [135].

Pugliese et al. conducted that was based upon resource dependency theory, focusing on the firm's profitability and board task performance current and previous efforts and utilizing responses from 264 CEOs from firms located in Italy [136]. The dependent variables are perceptual measures of board monitoring and advice tasks, with the independent variable being board task performance with two predictors firm past performance and industry regulation based upon data derived from archives and surveys [137]. A multivariate analysis was conducted to determine correlation and regression relationships. Post-hoc tests were performed to determine the robustness and validity of the proposed model [138]. The results of the study concluded that increased regulatory oversight produced complimentary board monitoring and advisement to CEOs, but doing so decreased the profitability of the organization [139]. This was supported by additional findings that suggested profitable firms have board members who are not as engaged with the CEO and further reinforces Ocasio's attention-based view of a firm's performance [140]. The CEO and its board members should be more actively engaged during the business model change from being dependent upon Chinese made products, creating a new internal model for American production and manufacturing process [141].

Rey and Powell focused on public universities and how declining revenue and increased costs present an opportunity to utilize resource dependency theory ("RDT") and determine strategies that can increase resource capacities [142]. The study was based upon multiple literature reviews and provided no quantitative or qualitative results, but did offer some potential methods to be further examined utilizing RDT [143]. What is apparent is that the model revolved around public higher education that seemed to be the dependent variable [144]. The independent variables of the model of public higher education were environmental impact, environmental constraints, problems for obtaining resources, and organizational effectiveness [145]. One of the research questions that could be applied to an RDT study is whether there a statistically significant relationship between public higher education and the independent variables' environmental impact, environmental constraints, problems for obtaining resources, and organizational effectiveness.

The proposed study could pose a more persuasive argument and provide results if conducted quantitatively due to the correlation among the variables. It could provide potential insights into resource strategies and offer suggestions to achieve specific goals. The research question posed above could be utilized to identify correlations and develop plans associated with resource strategies. One particular approach is leveraging social change, where universities partner with local communities to obtain access to not only locally deprived community resources but also federal funding [146-149]. This is a decisive business move given the potential for additional resources and federal funding, which is the lifeblood of most higher education institutions.

**The Implications for Social Change**

Dunfey defined social change as being the ability to change a current culture and social organization as time progresses, when creating an impact is vast and exponential such that its effects on society change the way we engage with others and how we create and sustain relationships [150]. For example, production increases in shale and fracking could open new markets and create clean energy products that could sustain the American way of life seemingly forever, while also providing fossil-fuel energy to other countries. Critics could argue that fracking hurts the environment. However, one could argue that depending on other nations to offer America goods, the absence of fracking harms the climate even more due to unfriendly environmental production processes in other countries [151]. Environmental campaigns could be implemented to encourage Americans from purchasing foreign product by educating individuals on the their current and future buying decisions [152].

TRA, TPB, and TAM incorporate numerous variables associated with measuring human behavioral factors. Researchers specializing in business could utilize these theories to identify processes to improve productivity and employee performance, as well as increasing other variables such as customer satisfaction, business expansion, appropriate technology implementation, and even create additional employment opportunities [153]. The potential social implications associated with TRA, TPB, and TAM could be reduced employee turnover, reduced unemployment due to the hiring of new personnel, improved products and delivery times, as well as organizational expansion into other markets [154]. This could lead to even further unemployment rates due to new factories being established, thus further growing organizations.

The United States has numerous environmental controls in place. America should build upon those processes and reduce fracking's impact on the environment. Additionally, if China is forced to reduce its production, it would, in essence, reduce its environmental impact. Finally, America could prevent additional viral incidents from happening. If they do happen again, we would be able to produce the much need masks, ventilators, and other life-saving drugs necessary to stop the virus in its tracks, without disrupting our economy. We have an opportunity to make a massive positive social change here by savings millions of lives, improving our environment, and creating a livelihood for millions of Americans and generations to come.

There are tendencies as well as commonalities to consider. The trends identified are supply chain management organizations all operate in very similar manners [155]. An additional trend is to focus on solutions utilizing traditional business practices such as risk management, process improvement, and forecasting, all of which are narrow-minded in that they rely on industry-standard methods [156]. There are also commonalities across all supply chain management processes. One commonality, even a tendency, is the lack of interpretation and incorporation of behavior into current processes within supply chain management organizations [157]. When methods are on the same level, there may be a conflict created. The conflict theory of culture ("CTC") deals with the creation of culture and the establishment of norms within an organization [158]. In most cases, these conflicts define socially acceptable practices in industries that are the basis for successful profits and production [159].

When considering the impact of social change, every organization operates similarly but leaves out behavioral intention as related to a particular operation or process [160]. This behavioral variable defines the topic of behavior and technology within global supply chain management organizations [161]. To integrate a new process or control mechanism within an organization as diverse as international supply chain management organizations, without taking into account behavioral intent further creates CTC factors, can hinder an organization's growth without the organization being able to identify the cause of the growth stoppage [162]. Organizations specializing in global supply chain management cannot be narrow-minded and focus on one particular process [163]. Companies should focus on all of their operations, but not in the traditional sense. TRA, TPB, and TAM can fill the gaps CTC creates [164].

**Conclusion**

The dependency on Chinese made products puts the United States in a substantially dangerous situation. The United States is already taking measures to avoid scarcity of certain supplies such as ventilators. However, it is not nearly what is required to ensure the economic safety of the American people. Additional measures must be taken and sustained, as well as further expanding internal supply chain management processes within the country. The United States should increase production internally to prevent future supply chain management shortages. If the nation does not stop depending on foreign products from China, the next Covid-19 situation may seriously cripple our country, where rioting and a complete lack of life-saving medications necessary to sustain our healthcare system would exist.